\documentstyle[11pt,aaspp4,twoside,flushrt]{article}
\def\2p{{$2^{nd}P$}}
\def\ldz{{$B-R/B+V+R$}}

\begin{document}                                                      
\noindent{\bf                                                         
HB MORPHOLOGY AND THE SECOND PARAMETER EFFECT: FAINT STARS IN A BIG GAME
}                                                                     

\begin{list}{}{\topsep 0in                                            
	  \partopsep 2\baselineskip                                   
	  \itemsep 0pt                                                
	  \parsep \baselineskip                                       
	  \leftmargin .53in                                           
	  \listparindent 0in                                          
	  \labelsep 0in                                               
	  \labelwidth 0in}                                            
\item ~                                                               


Flavio Fusi Pecci and Michele Bellazzini 

Osservatorio Astronomico, Bologna, and Stazione Astronomica, Cagliari, 
Italy

\end{list}                                                              


\vspace{2\baselineskip}                                                

\noindent{ABSTRACT:}                                                  
We schematically review some of the main steps in the study of
the Horizontal Branch Morphology in Galactic Globular Clusters.
Since the first realization of the existence of the so-called 
{\it Second Parameter Problem} (\2p) up to now, one could perhaps see
a sort of circular path. Actually, many candidate \2p's were proposed
during the early '70s, {\it age} has been {\it the} top-scored \2p 
during the late '80s, and we are now back to even more candidates,
including also possible intriguing combinations of some of them.


\vspace{\baselineskip}                                                

\pagestyle{myheadings}                                                
\markboth{\hspace*{1.0in}{\rm                                         
Fusi Pecci \& Bellazzini}\hspace{\fill}}
{{\rm                                                 
The Second Parameter }} 


\section{Introduction}
As a great movie star, the horizontal branch (HB) of  globular clusters has been 
on stage for more than four decades, playing an obvious major r\^ole in many 
astrophysical "big games", like the distance scale and the lower
limit to the age of the universe. Nevertheless, many fundametal aspects 
about origin and evolution of HB stars themselves are still wrapped up behind 
a thick smokescreen. Many valuable investigators have joined their efforts
to unveil such a curtain, finding a lot of important clues (sometimes
confusing, indeed), but very few clear-cut and solid conclusions have been
drawn on many relevant aspects so far. In particular, {\it firm} evidences
and conclusions adopted as {\it cornerstones} for some years, suddenly 
turned out to be {\it friable} or even misleading.

Despite their relative faintness, HB stars are implicated in several
astrophysical basic questions. Looking at a typical color-magnitude 
diagram (CMD), the comprehension of their distribution is fundamental 
both in the {\it vertical} sense, ({\it i.e.} luminosity: connected to 
distance and age scales, etc.) and in the {\it horizontal} sense,
({\it i.e.} temperature: connected to mass loss, integrated colors of 
distant stellar populations, etc.).

Here we briefly deal with this latter issue, {\it i.e.} the
HB morphology --the color (temperature) distribution of HB stars--
in Galactic globular clusters (GGCs), often also referred to 
as the {\it Second Parameter Problem} --(\2p). We provide a very 
schematic (and necessarily biased) reference table on about forty 
years of investigations, starting from the pioneering photographic works 
up to the latest studies carried out with HST, 
which leave us with the uncomfortable evidence that one of the most 
prominent features in the GGC CMDs is probably still one of the less 
understood phases of the evolution of Pop. II stars (see for instance 
the CMDs presented in these proceedings by Liebert et al. and 
Ferraro et al. showing HB's with 2-3 gaps).

To follow a sort of (admittedly biased) chronological path a parallel 
consultation of Table 1 may be useful, where just a few "reference" 
references can be found. We apologize for any unforgivable omission.

\section{Realizing that the HB-morphology world is not so simple...}
Though Shapley (1919) already had some suspicion, the realization of 
the existence of a \2p$-problem$ in the classification of the 
observed HB morphologies in GGCs came out in the early '60s.
In fact, the conclusion was: metallicity -- Z -- is the {\it first}
parameter; since however GGCs with similar metallicities have {\it different}
HBs, a {\it second} parameter must exist, which originates the
differences. 

Important efforts to compute the first HB theoretical models and to
yield a systematic parametrization of the observed HBs were made
during the early '70s. One can easily ascertain that the essence of 
the problem was already set on clear grounds at that time (see, in
particular, Rood 1973), and various \2p$-candidates$ were proposed
(helium abundance, age, CNO). Four items were especially noted, and 
frequently forgiven later:
\par\noindent
$\bullet$ {\sl The HB is not an evolutive sequence, but simply a narrow,
composite locus}, like {\it "a beach, where all stars go to take a bath"}.
\par\noindent
$\bullet$ {\sl Mass loss is the crucial phenomenon and parameter
which drives the actual location of the HB in color.} In particular,
{\it "why, when, where, how much, and how long mass loss does take
place before the HB matters very much"}.
\par\noindent
$\bullet$ {\sl The (color) temperature shift along the Zero Age HB
is strongly dependendent on mass loss and is highly non-linear
with varying metallicity}, $\Delta T_{eff}=f(\Delta M_{loss}, Z)$.
\par\noindent
$\bullet$ {\sl Any mechanism (intrinsic or induced) which could
somehow affect, during any evolutionary stage, the core and/or
total mass of the star may play a r\^ole in the \2p$-game$
because of the exceptional sensitivity of HB stars to any tiny variation.}

\subsection{Exotica enter into the game}
With the growing discoveries of peculiarities in the
chemical abundances of GGC stars and the existence of gaps and
special features along the main branches in the CMD (including the
HB), it became clear during the next decade ('73 -- '83) that the
interpretation of the HB morphology could not be so {\it simple}
as most of the players could hope. Other \2p's were proposed (in
particular, core rotation). It was also evident that most
of the observed chemical peculiarities could not be explained just by
{\it primordial} abundance variations, but that also special mixings
have to take place in the stars (see Kraft 1979, and Table 1).

At the end of that decade, in the attempt to explain the "curious"
couple of very metal-poor clusters M15--NGC5466, we (Buonanno et al. 
1985) naively proposed to ascribe their stricking different HB-morphologies 
to the very different {\it stellar density} within their central 
regions. Though hard to quantitatively estimate to what extent and how
such a new ingredient could play a significant r\^ole in the \2p$-problem$
(but surely via "induced" extra mass-loss), this idea built up a possible 
new bridge between stellar evolution and stellar dynamics.

\section{...and, somehow, forgetting it...}
In 1986, but actually starting from the fundamental papers by
Searle and Zinn (1978--SZ) and Zinn (1985), through the use of
the new index \ldz ~~~-- with $B,V,R=$number of blue, variable, red HB stars--
and of the overall description of the GGC system proposed by Zinn 
(1986),
two basic conclusions were achieved:
\par\noindent
$\bullet$ The \2p-effect presents a quite strong dependence on Galactocentric 
distance, with redder HB's predominantly found in the outer regions of
the Milky Way.
\par\noindent
$\bullet$ Since the most obvious way of producing redder HB's for fixed
metallicity is that of increasing the total mass (at constant core mass
and total mass loss), and since by decreasing the cluster age one 
automatically yields larger initial (TO) masses, the above evidence
has been {\it adopted} to be a clear-cut evidence that (a) {\sl age is}
{\it the} {\sl Second Parameter}, and (b) at given metallicity,
{\it redder HB's imply younger age}. Hence, the HB-morphology could
actually be used as a {\it clock}.

\noindent
While the interpretation of the {\it global} scenario has probably
been one of the most exciting and profitable steps in the description
of formation and evolution of our own Galaxy (and even of galaxies,
in general), there has also been a sort of "bouncing-back" effect 
leading to conclude that: since the adoption of age as {\it the only 
(dominant)} \2p allows us to build up a nice overall description of the 
Galaxy, then the \2p$-problem$ is solved. In our opinion, this was
--and still is-- not true.

\subsection{Global and non-global \2p's}
During the late '80s, the idea that the age is by far the {\it dominant}
parameter was corroborated by many observations (see Table 1). At the same 
time, however, also the detection of both photometric and spectroscopic 
peculiarities was growing up steadly, and the need for the distinction between 
the so-called {\it global} and {\it non-global} \2p's, proposed first by
Freeman and Norris (1981), became obvious though not unequivocal.
In fact, for instance Freeman and Norris assumed as {\it global}
the parameters which affect all the stars in a cluster at the same
extent; Lee (1993) defined as {\it global} the \2p which originates 
the dependence of the index \ldz ~~on Galactocentric distance;
we (Buonanno et al. 1985, Fusi Pecci 1987--$2^{nd} Conf. FBS$) 
interpreted the definition
of {\it global} to mean that the parameter driving the HB morphology is 
actually the result of the "global combination" of many individual
quantities and phenomena affecting the evolution of a single star
in a cluster or even the cluster as a whole.

\subsection{Age at the apex, but ....}
The scenario proposed first by Searle and Zinn (1978) has been 
very interestingly refined during the early '90s. And the idea that 
age is {\it the} \2p reached its apex in popularity.
Meanwhile, for the first time, {\it young} Galactic globular clusters have 
actually been found (see Table 1), using measurements based on the MS-TO.
Note that, due to the still too large uncertainties affecting all
age determinations (even {\it relative}), only for a very small 
sub-set of GGC's there is a sufficiently reliable evidence that 
they are actually {\it younger} than the bulk of clusters having
similar metallicities. 
Since these {\it young} globulars seem to be members of the Sgr
dwarf falling onto the Galaxy central regions, we are probably observing a
"real-time" confirmation of the {\it building block accretion}, as envisaged
by SZ.

But the questions are: (1) {\sl What 
fraction of the halo is actually young? How big is the age difference?
Where and how was this young component orignated?}, and (2) {\sl
Do the other globulars, presumably not involved in these "interacting
connections", actually show reliably significant age differences?}
Moreover, (3) {\sl Are we still allowed to include all the detected
special features just within the} {\it non-global} {\sl box?}
They are actually increasing in type and number, and "exceptions" are
becoming "the rule".

\section{... problems come back, ...}      
The results obtained in the latest years, especially thanks to
{\it HST}--observations in the very central regions of the clusters
and for Galactic globulars located in the inner bulge or at the
outskirts of the Milky Way, have shown that the interpretation of
the \2p--effect as due only to age cannot survive (see Table 1).
Detailed discussions on these new results and issues can be found in 
the recent reviews by VandenBerg et al. (1996), Stetson et al. (1996), 
Richer et al. (1996), Buonanno et al. (1997--AJ subm.).

\subsection{... and Exotica rise again}
Three pieces of evidence, at least, are in fact blowing up (again):
\par\noindent
$\bullet$ The HB-morphology of many clusters, especially if the very
central regions are observed, is very complex. Many clusters display
clearly bimodal distributions in color, there are clumps and gaps
(f.i. in NGC 1851, 2808, 6388, M 13, etc.), and the description of
the observed HB population and morphology with just one observational
parameter is clearly inadequate.

\par\noindent
$\bullet$ There is a growing evidence that significant populations of
binaries or binary descendants (primordial, collisional, mergers. etc.)
are present probably in almost any cluster. Moreover, internal radial
gradients in color, star population, and mass distribution testify the
importance of cluster  dynamical evolution, which may strongly affect also
the individual star evolution.

\par\noindent
$\bullet$ Coupled with many renewed efforts in model computation, the 
increasing pattern of information on chemical  abundances for many 
stars in several clusters obtained from high resolution
spectroscopy suggests (more and more convincingly) that mixing phenomena
actually take place in most stars in GGCs, and possibly at a different
extent from star-to-star. For instance, Sweigart (1996--preprint) has shown
that important variations in the helium abundance (a possible \2p) 
can be originated before the HB phase because of mixing, possibly
somehow coupled with rotation.

\section{Are we back to the 1973 status?}
It is hard to answer as we could say: Yes, after the latest evidences
we are back to play with too many \2p--candidates. On the other hand,
such a statement would be too strong as we now know that the HB-morphology
presents different properties and points out different features
depending on the observable adopted to describe it, as well as any 
astronomical object may look very different depending on the bandwidth 
adopted to observe it.  This implies that we must look at the 
HB-morphology and, in particular, at the \2p--problem within a
{\it global} approach, i.e. recalling that the HB-morphology is
in a sense {\it the final convolution of many different ingredients}.
We must first deconvolve the individual contributions before
ranking their importance in a given cluster.

{\footnotesize
{\it Due to shortage of space, appropriate references can be found 
in the few quoted papers in Tab. 1.}
}



\bigskip
\footnotesize
\begin{table}[h]
\begin{tabular}{|l|c|c|c|}
\multicolumn{3}{c}{\bf Table 1 - The 2nd Parameter Problem: 
some (admittedly biased) highlights}\\
\\
\hline
years     &         reference         &clue \\
\hline
          &                           &      \\
$'19$     & Shapley                   & M3 - M13 different RR's + HB's?\\
\hline
&&\\
$'52-'59$ & Arp, Baum, Kinman, Sandage & M3 - M92 different RR's + HB's\\
\hline
&&\\
$'60$     & Sandage, Wallerstein      & [Fe/H] is the first Parameter\\ 
          &                           & M13, M22 too blue for their metals \\
&&\\
$'65-'67$ & van den Bergh             & M3,M13, NGC 7006 etc. \\
          & Sandage, Wildey           & \2p: age? Helium?\\
&&\\
$'66-'69$ &Faulkner, Iben, Rood,      & first HB models; \2p: [CNO/Fe]?\\
          &Castellani, Giannone, Renzini &  mass loss is crucial\\
\hline
&&\\
$'72$     & Dickens            & HB-types: 1-7, non-monotonic with [Fe/H]\\
&&\\
$'72$     &Zinn                &spectr. peculiarities\\ 
          &                    & anomalous cepheids (V19 in NGC 5466)\\
&&\\
$'73$     & Mironov            & HB-types: B/B+R, a quantitative HB-index \\
\hline
&&\\
$'73$   & Rood, ApJ 184,815      &HB-syntethic models with mass loss\\
   &                             &\2p Candidates:\\ 
   &                             &Helium Abundance, Age, [CNO/Fe]\\
\hline
&&\\
$'73$   &Newell                   & HB gaps?? in the field? \\
\hline
\end{tabular}
\end{table}
\vfill\eject
\begin{table}[h]
\begin{tabular}{|l|c|c|c|}
\multicolumn{3}{c}{tab. 1 - continued} \\
\\
\hline
years    &  reference                         &clue \\
\hline
&&\\
$'73$   &Cannon, Lee              & NGC 6752: HB blue tail + gap \\
&&\\
$'74-'75$ &Harris                 & NGC 2808: bimodal HB + with wide gap\\
\hline
&&\\
$'75$   & Reimers                 &mass loss rate: basic formula\\
&&\\
$'76$   & Mengel, Gross           &models with (core) rotation\\
&&\\
$'77$   & Renzini                 & \2p Candidate: core rotation\\
\hline
&&\\
$'78-'80$ &Searle, Zinn, ApJ 225,357 &HB-Morphology varies systematically\\ 
   &                               & with Galactocentric distance\\
   &                               & \2p: $=$ Age \\
\hline
&&\\
$'79$   &Sweigart, Mengel          & meridional circulation $-->$ mixing\\
&&\\
$'79$   &Kraft, ARAA 17,309         & review: chemical peculiarities et al.\\
&&\\
$'79-'83$ &Norris, Smith, Pilachowski, & abundance: individual elements\\
   & Wallerstein, Kraft, Sneden,  & variations + correlations, etc.\\
   & Gratton, Bell, Kurucz, et al.   & models etc. \\
&&\\
$'79-'85$ &Cohen, Frogel, Pilachowski, & abundance scale: \\
   & Zinn, West, Peterson, Kraft, & wide and hot debate \\
   & Sneden, Gratton, Hesser, Bell, et al. & integr. phot.; high vs. low res.\\
$'81$   &Freeman, Norris, ARAA 19,319 & HB bimodality, CN, ellipticity, etc.\\
        &Suntzeff                    &  rotation, etc., peculiarities\\
        &                            &  \2p: "global" vs. "non-global"\\
\hline
&&\\
$'81$   &Grindlay, Bailyn, etal.      &HB-tail: (ex-)binaries ??\\
&&\\
$'81$   &Castellani, et al.           &HB-tail: RGB-manque' ??\\
&&\\
$'81-'83$ &Van Albada, De Boer, Dickens  &UV-properties: EB, B, I, R\\
          & Caloi, Castellani et al.     &                          \\
&&\\
$'81-'93$ & Sandage                  & Period-shift effect   + \\
          & Cacciari, Caputo, et al. & Oosterhoof dichotomy  etc.  \\ 
&&\\
$'83$  & Buzzoni et al.           & Helium: constant in Galactic GCs? \\
       &                          & (Iben 1968: R-method)\\
&&\\
$'83$ & Castellani, Renzini       &HB: non-monotonic behaviour\\
      &                           & $-->$ different metallicity regimes \\
\hline
\end{tabular}
\end{table}
\vfill\eject

\begin{table}[h]
\begin{tabular}{|l|c|c|c|}
\multicolumn{3}{c}{tab. 1 - continued} \\
\\
\hline
years    &  reference                         &clue \\
\hline
&&\\
$'83-'87$ &Buonanno,Corsi, FusiPecci  & M15 vs. NGC5466: $-->$ \2p Candidate:\\
   &  A\&A 145,97                     & environment (indirect)\\
   &                                  & + "global" combination\\
&&\\
$'83-'85$ &Peterson                 &rotation observed: \\
          &                         &M13 fast? M3, NGC 288 slow?\\
\hline
&&\\
$'85-'89$  &Iben, Tutukov, Bailyn et al. &BHB stars: from binary mergers??\\ 
&&\\
\hline
&&\\
$'85-'95$ &Zinn, ApJ 293, 424;  Armandroff,  & "global" vs. "non-global" \2p\\
  & Demarque, Lee, Sarajedini, & "global": depends on Galactocentric distance\\
   & Carney, King, Chaboyer et al.   & the only (dominant) global \2p: AGE\\
\hline
&&\\
$'88-'93$ &Djorgovski, King, Piotto, Bailyn, &Color/population gradients\\
          &Grindlay, Stetson, et al.         & Segregation? Interactions?\\
&&\\
$'88-'96$  &Crocker, Rood, O'Connell    &HB: Multiple populations??\\
         & Dorman, Renzini, Greggio, etal. &HB and UV flux etc. \\
&&\\
$'88-'92$ & Fusi Pecci, Renzini, Ferraro &Blue Straggler descendants on the 
                                   Red HB?? \\
&&\\
$'88-'96$ & Fusi Pecci, Bellazzini, Ferraro & Environment and blue HB tails\\
     & Buonanno, Corsi, Rood, et al.  & new HB observables + parametrization\\
\hline
&&\\
$'89-'93$ &Bolte, Green, Norris, Dickens &NGC 288/362: $-->$  \2p = AGE\\
          & VandenBerg, Bolte, Stetson,  &Age: "horizontal" method\\
          & Sarajedini, Demarque, King   &                        \\
\hline
&&\\
$'89-'94$ & Stetson, Ortolani, Gratton  & Pal 12: YOUNG based on MS-TO\\
          & Buonanno, Corsi, Ferraro, et al.
                              &Rup 106, Arp2, Ter 7, IC4499:\\
          & Richer, Fahlman   &YOUNG based on "vertical" method\\
\hline
&&\\
$'90-'94$ & Zinn, Armandroff, Lee &sub-systems in the GGC system:\\
   &  van den Bergh, Hesser,   & Young, Old; Halo, Disk, etc.\\
   &  Majewski, Freeman, & observables: kinematics, metallicity +\\
   &  Sarajedini, et al.      & clock: HB-morphology (2nd P = age) \\
&&\\
$'88-'95$  & Kraft et al.         & O, Na, Al, etc. peculiarities\\
$'93-'95$  & Rich, Liebert, Minniti & Metal rich GGCs: UV-excess\\
$'93-'95$  & Catelan, Freitas de Pacheco & NGC 288/362: revisited \\
           & Rood, Ferraro, et al. & HB(N2808) = HB(N362)+HB(N288)\\
           &                       & Age cannot be the "only" \2p\\
\hline
\end{tabular}
\end{table}
\vfill\eject

\begin{table}[h]
\begin{tabular}{|l|c|c|c|}
\multicolumn{3}{c}{tab. 1 - continued} \\
\\
\hline
years           &reference             &clue \\
\hline
&&\\
$'91$   &Preston, Shectman, Beers & "HB-morphology" for field stars  \\
        &Suntzeff, Kinman, Kraft, & RR Lyrae and HB-morphology \\
\hline
&&\\
$'92$     &Lin, Richer                  & YOUNG GGC's: captured ??\\
&&\\
$'94-'96$ & Ibata, Gilmore, Irwin &Sgr dSph merging with the Galaxy\\
         & Mateo, Sarajedini, Layden, &carrying its own GC system?\\
         &    et al.           &young globulars: Ter 7, Arp 2, etc.?\\
\hline
&&\\
$'95$   &Bailyn, ARAA 33,133 & review: Binaries in GGCs \\
        &                            & environment and stellar evolution\\
\hline
&&\\
$'94-'96$ &Ortolani, Renzini, Barbuy, etal. & HST Photometry of 
                                                 bulge GGCs:\\
          &                            & they are old, like 47 Tuc\\
$'95$ & Minniti  & bulge GGCs vs. disk GGCs\\
&&\\
$'94-'96$ &Richer et al. ApJ 463,602 & HST Photometry of remote halo GGCs: \\
        & Vandenberg et al. '96, ARAA, 34 & despite their red HB morphology,\\
        & Stetson et al. '96 PASP in press & they display the same age as \\
        & Bolte, et al.  & the inner halo GGCs of the same metallicity:\\ 
        &van den Bergh '96, PASP 108,986&                              \\
        & Buonanno et al. & age cannot be the dominant \2p $-->$ \\
        &                & the trend of HB-morph. with Galactocentric \\
        &        & distance cannot be straightforwardly \\
        &        & interpreted as an age gradient\\
&&\\
\hline
&&\\
$'96$   &Buonanno et al.  & dense environment and Blue Tails are correlated\\
        &                 & \\
\hline
&&\\
$'96$   & Sosin, Djorgovski, Piotto, King & HST photometry of bulge clusters:\\
        & Liebert, Rich, Dorman et al.& extended Blue Tail in metal-rich populations\\
&&\\
\hline
&&\\
$'96$   & Sweigart                  & New HB models: with extra-mixing\\
    &                               & \2p: Helium (+ rotation?) ?\\
&&\\
\hline
\hline
\end{tabular}
\end{table}



\end{document}